\documentclass[11pt]{amsart}   % For Latex2e
\usepackage{amssymb,xy}
\usepackage{amsthm, amsmath}
\usepackage[latin1]{inputenc}
\usepackage[T1]{fontenc}
\xyoption{all}
\theoremstyle{plain}
\theoremstyle{definition}
\theoremstyle{remark}
\title{Formulating problems for real algebraic geometry}
\author[Matthew England]{Matthew England}
\address{\small \rm  University of Bath, UK.}
\email{M.England@bath.ac.uk}

\usepackage{graphicx}

\begin{document}

\begin{abstract}
We discuss issues of problem formulation for algorithms in real algebraic geometry, focussing on quantifier elimination by cylindrical algebraic decomposition.  We recall how the variable ordering used can have a profound effect on both performance and output and summarise what may be done to assist with this choice.  We then survey other questions of problem formulation and algorithm optimisation that have become pertinent following advances in CAD theory, including both work that is already published and work that is currently underway.
With implementations now in reach of real world applications and new theory meaning algorithms are far more sensitive to the input, our thesis is that intelligently formulating problems for algorithms, and indeed choosing the correct algorithm variant for a problem, is key to improving the practical use of both quantifier elimination and symbolic real algebraic geometry in general.   
\end{abstract}

\maketitle

\section{Introduction}

\noindent We will discuss the effect problem formulation can have on the use of symbolic algorithms for real algebraic geometry.  This follows our recent work on cylindrical algebraic decomposition, one of the most important algorithms in this field.  We discuss the issue of variable ordering, well known to play a key role, but also survey a number of other issues that are now pertinent.  

Let $Q_i \in \{\exists,\forall\}$ and $\phi$ be some quantifier-free formula.  Then given
\[
\Phi := Q_{k+1}x_{k+1} \ldots Q_n x_n \, \phi(x_1,\ldots,x_n),
\]
\emph{quantifier elimination} (QE) is the problem of producing $\psi(x_1,\ldots,x_k)$, a quantifier-free formulae equivalent to $\Phi$.  In the case $k=0$ we have a \emph{decision problem}: is $\Phi$ true?  Tarski proved that QE is always possible for semi-algebraic formulae (polynomials and inequalities) over $\mathbb{R}$ \cite{Tarski98}.  The complexity of Tarski's method is indescribable as a finite tower of exponentials and so when Collins gave an alternative with cylindrical algebraic decomposition (CAD) \cite{Collins1975} it was a major breakthrough despite complexity doubly exponential in the number of variables.  CAD implementations remain the best option for many classes of problems.

Collins' CAD algorithm works in two stages.  First \emph{projection} calculates sets of projection polynomials $S_i$ in variables $(x_1, \dots, x_i)$ by applying an operator recursively starting with the polynomials from $\phi$.   Then in the \emph{lifting} stage decompositions of real space in increasing dimensions are formed from the roots of those polynomials.  First, the real line is decomposed according to the roots of the univariate polynomials.  Then over each cell $c$ in that decomposition the bivariate polynomials are taken at a sample point and a decomposition of $c \times \mathbb{R}$ is produced according to their roots.  Taking the union gives the decomposition of $\mathbb{R}^2$ and we proceed this way to a decomposition of $\mathbb{R}^n$.  The decompositions are cylindrical (projections of any two cells onto the first $k$ coordinates are either identical or disjoint) and each cell is a semi-algebraic set (described by polynomial relations).  

Collins' original algorithm uses a projection operator which guarantees CADs of $\mathbb{R}^n$ on which the polynomials in $\phi$ have constant sign, and thus $\Phi$ constant truth value, on each cell.  Hence only a sample point from each cell needs to be tested and the equivalent quantifier free formula $\psi$ can be generated from the semi-algebraic sets defining the cells in the CAD of $\mathbb{R}^k$ for which $\Phi$ is true.  There have been numerous improvements, optimisations and extensions of CAD since Collins' work (with a summary of the first 20 years given in \cite{Collins1998}).

\section{Variable ordering}

When using CAD for QE we must project quantified variables first, but we are free to project the other variables in any order (and to change the order within quantifier blocks).  The variable ordering used can make a big difference.  For example, let $f:=(x-1)(y^2+1)-1$ and consider the two minimal CADs visualised below.  In each case we project down with the left figure projecting $x$ first and the right $y$.  In this case we see that wrong choice more than doubles the number of cells.  
Of course, this is just a toy example, but \cite{BD07} defined a class of examples where changing variable ordering would change the number of cells required from constant to doubly exponential in the number of variables.
\begin{center}
\includegraphics[width=0.45\textwidth]{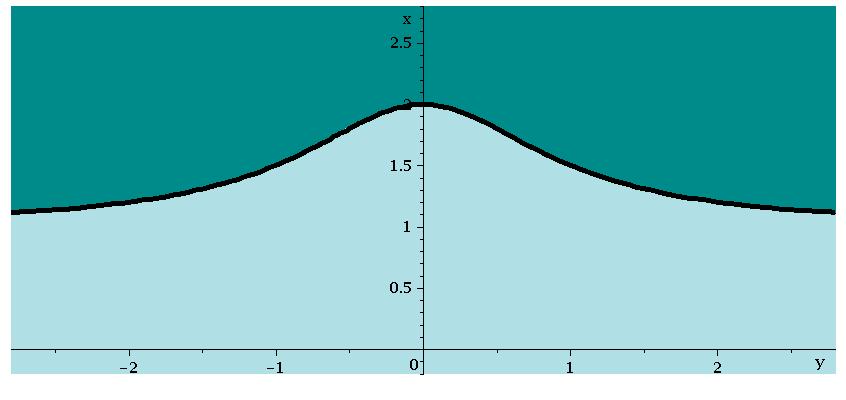}
\hspace*{0.05\textwidth}
\includegraphics[width=0.45\textwidth]{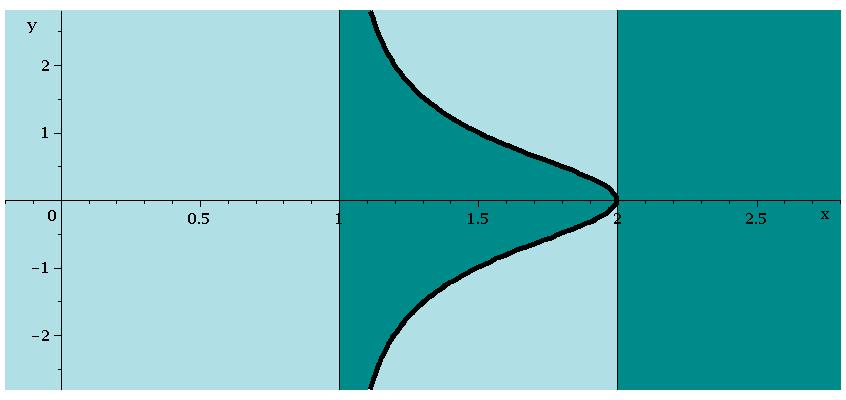}
\end{center}
Various heuristics exist to help choose a good variable ordering:
\begin{description}
\item[Brown] Eliminate lowest degree variable first (with tie-breaking rules) \cite[Section 5.2]{Brown2004}.  Quite effective but considers only the initial input rather than the full projection set.
\item[sotd] For all admissible orderings, calculate the projection set and choose the one with smallest \emph{sum of total degree} \cite{DSS04}.  Performs well but costly with many orderings.  
\item[Greedy sotd] Allocate one variable of the ordering at a time by projecting each unallocated variable and choosing the one which increases the sotd least \cite{DSS04}.
\item[ndrr] The sotd based heuristics can be misled, or give ties, especially when the differences lie in the real geometry.  In this case we can compare the \emph{number of distinct real roots} of the univariate projection polynomials \cite{BDEW13} (the first step of lifting).
\item[Machine Learning] Essentially a meta-heuristic on the above \cite{HEWDPB14}.
\end{description}

\section{Other questions of input formulation}

A key improvement to CAD is the development of projection operators that guarantee only truth invariance of $\phi$, rather than sign-invariance of the polynomials within.  This is achieved by considering the logical structure of $\phi$, but brings in sensitivity to such structure.
\begin{description}
\item[Designating ECs]  An \emph{equational constraint} (EC) is an equation logically implied by a formula.  The algorithm in \cite{McCallum1999} builds a CAD relative to a designated EC which is sign-invariant for the polynomial defining the EC, and for the other polynomials only when the EC is satisfied. 
If a formula has more than one EC, which to designate?
\item[Sub-formulae for TTICAD]  In \cite{BDEMW13} a truth-table invariant CAD (TTICAD) was defined as a CAD on whose cells the truth-table for a set of formulae is invariant.  A new operator was presented which takes advantage of ECs in the separate formulae.  If any formula has more than one EC then we have the issue above again.  Further, TTICAD can be used to find a truth-invariant CAD for a single formula by breaking it up into sub-formulae, but how best to do this?
\end{description}
Experimental results in \cite{BDEMW13} suggested the heuristics above can also help with these questions, but when the issues are combined the number of possibilities can become overwhelming.  Further, there are additional issues where the existing heuristics are of no help.
\begin{description}
\item[Order to process constraints]  In \cite{BCDEMW14} a new TTICAD algorithm is presented which is sensitive to the order in which constraints are considered.  The images below represent two TTICADs relative to a formula defined by the polynomials graphed.  The difference is caused solely by this ordering with the one on the right having three times more cells. In \cite{EBCDMW14} new heuristics are developed to help with this choice.
\item[Implicit ECs]  Consider 
$\phi := (f_1=0 \land \phi_1) \lor (f_2=0 \land \phi_2)$.  There is no explicit EC but the formula is logically equal to $(f_1f_2=0) \land \phi$.  Using this gives the benefit of the reduced projection set and thus less cells, but the increase in polynomial degrees may have an impact on timings.  
\item[Well-orientedness]  Some CAD algorithms only work on input that is \emph{well-oriented}.  The precise details of this condition varies between algorithms and it is possible for input to be well-oriented for one but not another.  This raises the question of whether a good choice can be made at the start, or if partial calculations can be reused?
\end{description}

\begin{center}
\includegraphics[width=0.45\textwidth]{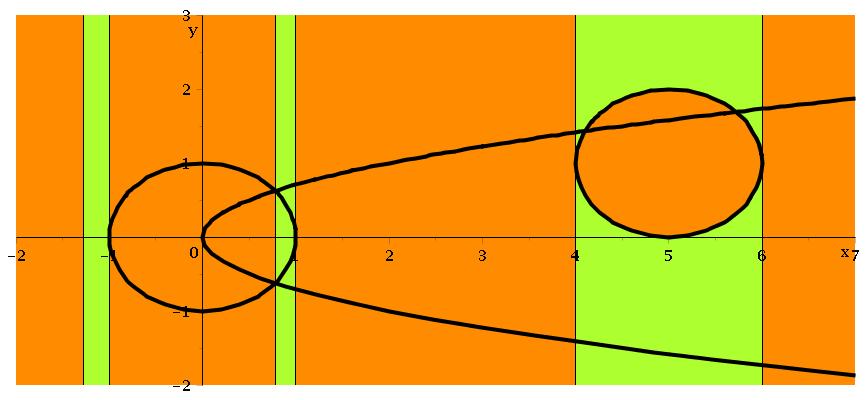}
\hspace*{0.03\textwidth}
\includegraphics[width=0.45\textwidth]{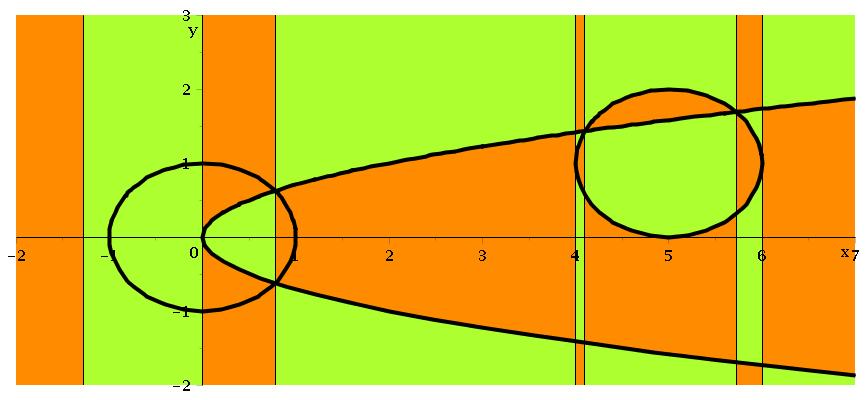}
\end{center}

\section{Formulating problems, preprocessing and algorithm choice}

Finally, we remark on some related issues which have come to light recently.
\begin{description}
\item[Precondition input]  In \cite{WBD12_GB} the idea of preconditioning the input to CAD using Groebner bases was investigated, with \cite{BDEMW13} extending this to TTICAD.  The former found that this could be extremely beneficial, but not universally so.  A heuristic was developed to identify when, but \cite{BDEMW13} found this was not suitable for TTICAD.
\item[Deriving the mathematics]  In \cite{WDEB13} a long standing motion planning problem was solved using CAD by changing the analysis used to formulate the input formula.  Instead of a description of the feasible region a negation of one for the infeasible region was used.  Such a reformulation was easy to do but made a great difference to the feasibility of CAD.  How can we identify such benefits in general?
\item[Algorithm choice]  
Recently an alternative to the projection and lifting approach to CAD has been investigated, in which  a decomposition of complex space is first built using triangular decompositions and regular chains theory \cite{CM12b} (this was how the TTICAD algorithm in \cite{BCDEMW14} differed from \cite{BDEMW13}).  
Experiments in \cite{CM12b} and \cite{BCDEMW14} show that the different approaches outperform each other for different examples.  How can we classify examples for use with one approach or the other?  
\end{description}

\section{Conclusions}

We have summarised issues of problem formulation which can dramatically affect the performance of CAD.  In some cases heuristics have been developed to help, but there is still much work to be done in making these practical and in extending them to the currently unanswered questions.  It is likely that much of what is learned here could be used throughout QE, or more generally for symbolic algebraic geometry.

\end{document}